%% file: art2.tex
\newcommand{\bea}{\begin{eqnarray}}
\newcommand{\eea}{\end{eqnarray}}
\newcommand{\be}{\begin{equation}}
\newcommand{\ee}{\end{equation}}
\newcommand{\ben}{\begin{enumerate}}
\newcommand{\een}{\end{enumerate}}

\def\eck#1{\left\lbrack #1 \right\rbrack}

\def\rund#1{\left( #1 \right)}
\def\abs#1{\left\vert #1 \right\vert}

\def\ave#1{\left\langle #1 \right\rangle}

\def\A{{\cal A}}
\def\B{{\cal B}}

\def\R{{\cal R}}

\def\hb{{\hfill\break}}

\def\eps{{\epsilon}}

\def\vp{\varphi}

\def\Real{{\rm I\mathchoice{\kern-0.70mm}{\kern-0.70mm}{\kern-0.65mm}%
  {\kern-0.50mm}R}}
\def\C{\rm C\kern-.42em\vrule width.03em height.58em depth-.02em
       \kern.4em}
\font \bolditalics = cmmib10
\def\bx#1{\leavevmode\thinspace\hbox{\vrule\vtop{\vbox{\hrule\kern1pt
        \hbox{\vphantom{\tt/}\thinspace{\bf#1}\thinspace}}
      \kern1pt\hrule}\vrule}\thinspace}

\def \vc #1{{\textfont1=\bolditalics \hbox{$\bf#1$}}}
{\catcode`\@=11
\gdef\SchlangeUnter#1#2{\lower2pt\vbox{\baselineskip 0pt \lineskip0pt
  \ialign{$\m@th#1\hfil##\hfil$\crcr#2\crcr\sim\crcr}}}
}

\def\ueber#1#2{{\setbox0=\hbox{$#1$}%
  \setbox1=\hbox to\wd0{\hss$\scriptscriptstyle #2$\hss}%
  \offinterlineskip
  \vbox{\box1\kern0.4mm\box0}}{}}

\def\bx#1{\leavevmode\thinspace\hbox{\vrule\vtop{\vbox{\hrule\kern1pt
        \hbox{\vphantom{\tt/}\thinspace{\bf#1}\thinspace}}
      \kern1pt\hrule}\vrule}\thinspace}

\def\elabel#1{\label{eq:#1}}

{\catcode`\@=11
\gdef\SchlangeUnter#1#2{\lower2pt\vbox{\baselineskip 0pt \lineskip0pt
  \ialign{$\m@th#1\hfil##\hfil$\crcr#2\crcr\sim\crcr}}}
}

\documentclass[onecolumn]{aa}
\usepackage{graphicx}
\usepackage[tbtags,fleqn]{amsmath}
\usepackage{pstricks,pst-node}
\usepackage{natbib}

\newcommand{\mincir}{\raise
  -2.truept\hbox{\rlap{\hbox{$\sim$}}\raise5.truept \hbox{$<$}\ }}
\newcommand{\magcir}{\raise
  -2.truept\hbox{\rlap{\hbox{$\sim$}}\raise5.truept \hbox{$>$}\ }}

\begin{document}
   \title{The three-point correlation function of cosmic shear: I. The
   natural components}
   \titlerunning{The three-point correlation function of cosmic shear}
   \author{
  Peter Schneider
          \inst{1,2}
          \and
          Marco Lombardi\inst{1}
          }

   \offprints{P. Schneider}

   \institute{Institut f\"ur Astrophysik und Extraterrestrische Forschung, Universit\"at Bonn,
              Auf dem H\"ugel 71, D-53121 Bonn, Germany\\
              \email{peter@astro.uni-bonn.de, lombardi@astro.uni-bonn.de}
         \and
                Max-Planck-Institut f\"ur Astrophysik, Postfach 1317,
              D-85741 Garching, Germany}

   \date{Received ; accepted }

   \abstract{The three-point correlation function of cosmic shear, the
   weak distortion of the images of distant galaxies by the
   gravitational field of the inhomogeneous matter distribution in the
   Universe, is studied here. Previous work on three-point statistics
   of cosmic shear has mainly concentrated on the convergence, or on
   aperture measures of the shear. However, as has become clear
   recently for the two-point statistics of cosmic shear, the basic
   quantity that should be used is the correlation function: first, it
   is much easier to measure from observational data, since it is
   immune against complicated geometries of data fields (which contain
   gaps and holes, e.g. due to masking); second, all other (linear)
   two-point statistics can be expressed as integrals over the
   correlation function. The situation is the same for the three-point
   statistics. However, in contrast to the two-point correlation
   function, the invariants (with respect to rotations) of the shear
   three-point correlation function have not been employed yet. Here
   we consider the transformation properties of the shear three-point
   correlation function under rotations. We show that there are four
   complex linear combinations of components of the three-point
   correlation function, which we shall call `natural components',
   since they are multiplied just by a phase factor for arbitrary
   rotations, but do not mix. In particular, their moduli are
   invariant under rotations and thus (non-linear) invariants of the
   three-point correlation function.  In terms of these natural
   components, the invariance of the statistical properties of the
   shear field under parity transformations are easily obtained. Our
   results do not apply only to cosmic shear, but also to other
   quantities with the same mathematical properties -- that of a
   polar. For example, practically every relation derived here applies
   also to the polarization of the cosmic microwave background
   radiation.

   }

   \maketitle
%

\section{Introduction}
The weak gravitational lensing effect by the large-scale matter
distribution of the Universe, called cosmic shear, has long been
recognized as a unique tool to study the statistical properties of the
cosmological (dark) matter distribution, without referring to luminous
tracers of this distribution (Blandford et al. 1991; Miralda-Escude
1991, Kaiser 1992, 1998, Jain \& Seljak 1997, Bernardeau et al.\ 1997,
Schneider et al.\ 1998, van Waerbeke et al. 1999; Bartelmann \&
Schneider 1999; Jain et al.\ 2000, White \& Hu 2000; see Mellier 1999
and Bartelmann \& Schneider 2001 for recent reviews).  Owing to the
smallness of the effect, its actual measurement has only fairly
recently been achieved, nearly simultaneously by several groups (Bacon
et al. 2000; Kaiser et al. 2000; van Waerbeke et al. 2000; Wittman et
al. 2000). This breakthrough became possible due to the usage of
wide-field optical cameras and the development of special-purpose
image analysis software specifically designed to measure the shape of
very faint galaxies and to correct their ellipticity for effects of
PSF smearing and anisotropy. By now, several additional cosmic shear
measurements have been reported (Maoli et al. 2001; van Waerbeke et
al. 2001; Rhodes et al. 2001; Bacon et al.\ 2002; Refregier et al.\
2002; H\"ammerle et al. 2002; Hoekstra et al.\ 2002), both from the
ground and from HST imaging, partly with appreciably larger sky area
than the original discovery papers.

In all of these papers, the cosmic shear signal detected was one
related to the two-point correlation function of the shear, or some
function of it, such as the shear dispersion or the aperture mass. To
measure higher-order statistical properties of the cosmic shear, the
quantity of data must be larger than for the second-order measures. It
has been pointed out by a number of authors (e.g., Bernardeau et al.\
1997; Jain \& Seljak 1997; Schneider et al.\ 1998; van Waerbeke et
al.\ 1999; Hamana et al.\ 2002) that the third-order statistics
(e.g. the skewness) contains very valuable cosmological information,
such as the density parameter $\Omega_{\rm m}$. In particular, the
near-degeneracy between $\sigma_8$ and $\Omega_{\rm m}$ in two-point
cosmic shear statistics (see, e.g., van Waerbeke et al.\ 2002)
can be broken if the three-point statistics is
employed.  Encouragingly, Bernardeau et al.\ (2002a) have
reported the detection of a third-order statistical signal in their
cosmic shear survey.

Apart from the larger difficulty to obtain a measurement of the
third-order statistics of the cosmic shear, there is also the problem
of an appropriate statistical estimator for the third-order
shear. Whereas for the second-order, the statistically independent
shear measures are known, we are not in this position for the
third-order shear statistics. We shall briefly summarize the situation
for the two-point statistics, and explain why the three-point shear
statistics is substantially more complicated in Sect.\ 2 below.  In
Sect.\ts 3 we shall then define the components of the shear
three-point correlation function (3PCF) and study their transformation
behavior under spatial rotations.  From that, we shall then find in
Sect.\ts 4 the natural components of the shear 3PCF, which can be
considered analogous to the natural components of the two-point
correlation function of the shear. These components are `natural' in
the sense that they have the simplest behavior under rotation
transformations: each of the four complex natural components is just
multiplied with a phase factor when an arbitrary rotation is applied,
which in particular means that the moduli of these natural components
are invariants under rotations.  We shall discuss the importance of
these natural components in Sect.\ts 5, where we also outline the
perspectives of future work that can be based on the use of these
natural components.  In an appendix, we shall consider the projection
of the shear onto several particular reference points, defined by the
various centers of a triangle.

We want to point out that all the relations derived in this papers are
not confined only to cosmic shear. In fact, this paper investigates
the three-point correlation function of a polar -- a polar is a
two-component quantity which transforms under a rotation of the
coordinate frame by a phase factor ${\rm e}^{2{\rm
i}\vp}$. Alternatively, a polar can be viewed as the trace-free part
of a symmetric $2\times 2$ matrix. In the case of cosmic shear, this
matrix is the Hessian of the deflection potential. Another polar of
great cosmological importance is the polarization of the cosmic
microwave background. Hence, all the results presented below do
equally well apply to the three-point correlation function of the CMB
polarization.

\section{Motivation}
For the two-point cosmic shear statistics, the basic quantities are
the two-point correlation functions. Given a pair of points, $\vc
X_i$, and the Cartesian components of the shear $\gamma_\mu(\vc X_i)$
there ($i=1,2$, $\mu=1,2$), one projects the shear along the direction
$\vp$ connecting these two points by defining the tangential and cross
component, $\gamma_{\rm t}$ and $\gamma_\times$ by $\gamma_{{\rm t}i}
+ {\rm i}\gamma_{\times i} = -[\gamma_1(\vc X_i)+{\rm i}\gamma_2(\vc
X_i)]\,{\rm e}^{-2{\rm i}\vp}$. Then, one forms the correlation
functions $\xi_{\rm tt}(\theta)=\ave{\gamma_{{\rm t}1}\gamma_{{\rm
t}2}}$ and $\xi_{\times\times}(\theta)=\ave{\gamma_{\times
1}\gamma_{\times 2}}$, where the average is an ensemble average over
all pairs of points with separation $\theta$. Even more useful are the
linear combinations $\xi_\pm(\theta)=\xi_{\rm tt}(\theta)
\pm\xi_{\times\times}(\theta)$.  Another combination which one may be
tempted to take is $\ave{\gamma_{{\rm t}1}\gamma_{\times 2}}$, but
this changes sign under parity transformations and thus should
vanish. All other two-point statistical measures of the cosmic shear,
such as the shear dispersion in a circle or the aperture mass
dispersion, can be expressed as integrals over these two correlation
functions (e.g., Crittenden et al.\ 2002; Schneider et al.\ 2002a). From
a practical point of view, the determination of the shear correlation
functions is also most convenient, as they can be measured also in
data fields of complicated geometry (as normally data fields are, due
to masking). The two correlation functions $\xi_\pm$ can be expressed
readily in terms of the power spectrum of the mass distribution in the
Universe.

Compared with this situation, a proper measure of the three-point
shear statistics is much more difficult to define. First we note that the
two-point function $\xi_+=\ave{\gamma \gamma^*}$ (where `*' denotes
complex conjugation) can be defined without any reference direction;
this is not the case for any three-point function of the shear, since
with three two-component quantities alone, no tri-linear scalar can be
formed. Hence, one needs to project the shear components. In contrast
to the case of two points, where there is a unique choice of the
reference direction, this is no longer true for a triangle: there is
not a single `natural direction' defined in a triangle; in fact, there
are several of those (see the Appendix). Therefore, it is not a priory clear
how to define `useful' components of the three-point correlation
function. One might ask, for example, whether there are similar
`invariant' combinations of the components of the  shear 3PCF
as there are for the two-point function
($\xi_\pm$).

Given these difficulties, it is not surprising that the work on the
three-point statistics of cosmic shear has been relatively
sparse. Bernardeau et al.\ (1997) and van Waerbeke et al.\ (1999)
consider the 3PCF of the surface mass density as reconstructed from
the shear measurements. Whereas possible in principle, the fact that
real data sets have gaps and holes makes the reconstructed mass map
susceptible to systematics due to the geometry. Schneider et al.\
(1998) suggested to use the aperture mass (Schneider 1996) as a cosmic
shear statistics for which the third-order moment is readily
calculated directly from the shear data. However, as is the case for
the shear dispersion, one needs to cover the data field with
(circular) apertures which presents again a problem in case of gaps in
the data. Recognizing this, Bernardeau et al. (2002b) defined a
particular component (or, more precisely, a particular linear
combination of components) of the 3PCF that is readily measured from
observational data, calculated its expectation value from numerical
ray-tracing simulations of Jain et al.\ (2000) and successfully
applied it to the VIRMOS-DESCART survey in Bernardeau et al.\ (2002a).

In addition to the component of the three-point shear correlator considered
by Bernardeau (2002b), the other components (there are a total of 8)
may contain equally valuable information about the bispectrum of the
mass distribution. In addition, it is easily seen that all measures of
the third-order shear statistics can be expressed as integrals over
the shear 3PCF. They are easiest to determine
from real data and shall therefore be considered as the basic
quantities. In this paper we will derive the `natural' components of
the shear 3PCF, by considering their transformation behavior under
rotations. 

\begin{figure}[!t]
  \parbox[t]{0.49\hsize}{%
    \resizebox{\hsize}{!}{\input fig1.tex}
    \caption{Definitions of the geometry of a triangle. The $\vp_l$
    are the orientations of the sides of the triangle relative to the
    positive $x_1$-direction, the $\phi_l$ are the interior angles of
    the triangle, which are related to the $\vp_l$ by
    (\ref{eq:angles}), provided the orientation of the three points is
    as displayed here.}%
    \label{fig:1}}
  \hfill
  \parbox[t]{0.49\hsize}{%
    \resizebox{\hsize}{!}{\input fig4.tex}
    \caption{The orthocenter of a triangle, determined by the
    intersection of the three altitudes. The angle between the
    altitude of $\vc x_1$ and the side $\vc x_3$ is easily obtained
    from the right triangle formed by this altitude and the two sides
    $\vc x_1$ and $\vc x_3$.}%
    \label{fig:2}}
\end{figure}

\section{The shear three-point correlation function, and the centers
    of a triangle}
\subsection{Definition of the 3PCF}
Consider three points $\vc X_l$, $1\le l\le 3$, and define their
difference vectors $\vc x_1=\vc X_3-\vc X_2$, $\vc x_2=\vc X_1-\vc
X_3$, $\vc x_3=\vc X_2-\vc X_1$, so that $\vc x_1+\vc x_2+\vc x_3=\vc
0$ (see Fig.\ts\ref{fig:1}).  Each of the three difference vectors
will be written as $\vc x_l=(x_l \cos \vp_l,x_l\sin \vp_l)$, so that
$\vp_l$ is the orientation of the $l$-th side of the
triangle. Furthermore, we define $\phi_l$ as being the interior angle
of the triangle at the corner $\vc X_l$.  In order to have general
relations, in this paper we will use \textit{oriented angles}, i.e.\
we will attach to each angle a sign indicating its orientation.  More
precisely, we will define $\phi_1\in(-\pi,+\pi)$ to have the same sign
of the cross product $\vc x_2 \times
\vc x_3$ (where $\vc a\times \vc b:=a_1 b_2 -a_2 b_1$), 
and similarly for $\phi_2$ and $\phi_3$.  Note that, since $\vc x_1
\times \vc x_2 = \vc x_2 \times \vc x_3 = \vc x_3 \times \vc x_1$
(which follows from the vanishing of the sum of the $\vc x_l$, or in a
more geometric way, from the fact that each of these cross products
equals twice the area of the triangle), all angles $\phi_l$ will be
either positive or negative; in particular, they all will be positive
if the closed path from $\vc X_1$ to $\vc X_2$ to $\vc X_3$ to $\vc
X_1$ goes around the triangle counter-clockwise.  We also observe that
this convention for the angles implies that the sum of the internal
angles of the triangle will be $\pm \pi$ depending on the orientation;
however, this ambiguity will not generally play a role, since all
relations for angles are defined modulo $2 \pi$.  In the following we
will call a triangle positively (respectively, negatively) oriented if
the sum of its internal angle is $+\pi$ ($-\pi$).
The relation between the $\phi_l$ and the $\vp_l$ is given by 
\begin{align}
  \vp_3 - \vp_2 = {} & \pi - \phi_1\; ; &
  \vp_1 - \vp_3 = {} & \pi - \phi_2\; ; &
  \vp_2 - \vp_1 = {} & \pi - \phi_3\; .  
  \elabel{angles}
\end{align}

Let $\gamma_\mu(\vc X_l)$ be the Cartesian components of the shear at
point $\vc X_l$; they are defined in terms of the deflection potential
$\psi(\vc X)$ by the differential operation
$\gamma_1=[\partial^2\psi/(\partial X_1^2)-\partial^2\psi/(\partial
X_2^2)]/2$, $\gamma_2= \partial^2\psi/(\partial X_1\,\partial X_2)$.
We define the Cartesian components of the shear 3PCF as
\be
\gamma_{\mu\nu\lambda}(\vc x_1,\vc x_2,\vc x_3)\equiv
\ave{\gamma_\mu(\vc X_1)\,\gamma_\nu(\vc X_2)\,\gamma_\lambda(\vc
X_3)}\;,
\ee
where we have made use of the fact that the shear field is assumed to
be a homogeneous random field, so that the 3PCF is invariant under
translations; therefore, $\gamma_{\mu\nu\lambda}$ does depend only on
the separation vectors $\vc x_l$. Although we also assume that the
random field is isotropic, the Cartesian components
$\gamma_{\mu\nu\lambda}$ of the 3PCF do not depend just on the
$x_l=\abs{\vc x_l}$ (see below).

For any reference direction $\zeta_l$, we can define the
tangential and cross components of the shear, $\gamma_{l\rm t}$ and
$\gamma_{l\times}$, respectively, at point $\vc X_l$ relative to this
direction by
\be
\gamma(\vc X_l,\zeta_l)\equiv \gamma_{\rm t}(\vc X_l,\zeta_l)+{\rm
i}\gamma_{\times}(\vc X_l,\zeta_l) :=-\gamma(\vc X_l) {\rm e}^{-2{\rm
i}\zeta_l}=-\eck{\gamma_1(\vc X_l)+{\rm i}\gamma_2(\vc X_l)}
{\rm e}^{-2{\rm i}\zeta_l}
\;,
\ee
or explicitly in terms of the components,
\be
\gamma_\mu(\vc X_l,\zeta_l)=-R_{\mu\nu}(2\zeta_l)\gamma_\nu(\vc
X_l)\;,
\elabel{transf}
\ee
where $R_{\mu\nu}(\vp)=\delta_{\mu\nu}\cos\vp+\eps_{\mu\nu}\sin\vp$ is
the rotation matrix, $\delta_{\mu\nu}$ the Kronecker delta, and
$\eps_{12}=1=-\eps_{21}$, $\eps_{11}=0=\eps_{22}$; here and in the
following, we identify the tangential component with $\mu=1$, and the
cross component with $\mu=2$ when we write transformation equations
like (\ref{eq:transf}), and we employ the Einstein summation
convention.  We notice that, because of the rotation behavior of the
shear, we only need to specify the projection direction $\zeta_l$
modulo $\pi$.

If the reference directions $\zeta_l$ are defined in terms of the
position vectors of the vertices of the triangle, or the side vectors,
and thus rotate in the same way as the triangle as whole, then the
tangential and cross components of the shear are invariant under
rotations and translations of the triangle. For example, the direction
$\zeta_l$ could be chosen as the direction $\vp_l$ of the opposite
side of the triangle. The 3PCF of these projected shear components
will then depend only on the $\abs{\vc x_l}$,
\be
\gamma_{\mu\nu\lambda}^{(\zeta_l)}(x_1,x_2,x_3)
=-R_{\mu\alpha}(2\zeta_1)\,R_{\nu\beta}(2\zeta_2)\,R_{\lambda\gamma}(2\zeta_3) 
\gamma_{\alpha\beta\gamma}(\vc x_1,\vc x_2,\vc x_3)\;.
\elabel{transfo}
\ee
The orientation of the triangle becomes important if the arguments of
the 3PCF are written as the three side lengths; in this case, the
triangle is defined only up to a parity transformation.   Hence, in
all relations involving only the side lengths of a triangle, we will
assume that the triangle has positive parity.  We shall see
in Sect.\ts 4 how the shear 3PCF behaves under parity transformations.

In Eq.\ (\ref{eq:transfo}), the directions $\zeta_l$ are arbitrary. For a
given set of three points, i.e., for a given triangle, there are
several natural choices for the reference directions; we shall discuss
those in the next subsection.  One choice was already mentioned above, namely
the direction $\vp_l$ of the side opposite to the corner $\vc X_l$. We
shall label the corresponding 3PCF with the superscript `s' (for
`side'), 
\be
\gamma_{\mu\nu\lambda}^{(\rm s)}(x_1,x_2,x_3)
=-R_{\mu\alpha}(2\vp_1)\,R_{\nu\beta}(2\vp_2)\,R_{\lambda\gamma}(2\vp_3) 
\gamma_{\alpha\beta\gamma}(\vc x_1,\vc x_2,\vc x_3)\;.
\elabel{gamside}
\ee

\subsection{The centers of a triangle, and the corresponding shear
      projections} 
In addition to the side projection mentioned above,
there are several other natural choices for the directions along which
the shear can be projected. For each triangle, one can define a number
of `centers'; the direction of the vector connecting the point $\vc
X_l$ with one of these centers can be used to define convenient
components of the 3PCF.  The four most important centers are: (1) The
centroid of a triangle; it is the point where the side-bisectors
intersect; (2) the incenter (center of the incircle), which is the
intersection of the three angle-bisectors of the interior angles
$\phi_l$; (3) the circumcenter (center of the circumcircle), which is
the point of intersection of the three midperpendiculars, and (4) the
orthocenter, which is the intersection point of the altitudes.

\begin{figure}[!t]
  \parbox[t]{0.49\hsize}{%
    \resizebox{\hsize}{!}{\input fig2.tex}
    \caption{The incenter of a triangle is given by the intersection
    of the three interior angle-bisectors.
}%
    \label{fig:3}}
  \hfill
  \parbox[t]{0.49\hsize}{%
    \resizebox{\hsize}{!}{\input fig5.tex}
    \caption{The centroid of a triangle is the intersection point of
    the three side-bisectors.
}%
    \label{fig:4}}
\end{figure}

We now define the tangential and cross components of the shear
relative to the direction of the line connecting the point $\vc X_l$
with one of these centers. Since the line connecting $\vc X_l$ with
the orthocenter (the point `H' in Fig.\ts\ref{fig:2}) is perpendicular
to the side vector $\vc x_l$, we find for the projection of the shear
relative to the direction of the orthocenter (labeled with superscript
`o') simply by setting $\zeta_l=\vp_l+\pi/2$ in (\ref{eq:transf}),
\be
\gamma^{(\rm o)}_\mu=-\gamma^{(\rm s)}_\mu \;.
\elabel{o_to_s}
\ee
Therefore,
we obtain the shear 3PCF for this projection
from Eq.\ (\ref{eq:transfo})
\be
\gamma_{\mu\nu\lambda}^{(\rm o)}(x_1,x_2,x_3)=-
\gamma_{\mu\nu\lambda}^{(\rm s)}(x_1,x_2,x_3) \;.
\elabel{sidetoortho}
\ee
Next we consider the center of the incircle. As this is given by the
intersection of the angle-bisectors, we obtain for the direction
$\zeta_1$ of the line connecting $\vc X_1$ with the incenter (see
Fig.\ts \ref{fig:3}) $\zeta_1=\vp_3+\phi_1/2$; hence,
\[
\gamma_\mu^{(\rm in)}(\vc
X_1)=-R_{\mu\nu}(2\vp_3+\phi_1)\gamma_\nu(\vc X_1)\;.
\]
Using Eqs.\ (\ref{eq:transf}) and (\ref{eq:o_to_s}), one finds that the
relation between the Cartesian components of the shear and those
defined with respect to the ortho-center is
\be
\gamma_\mu(\vc X_l)=R_{\mu\nu}(-2\vp_l)\gamma_\nu^{(\rm o)}(\vc X_l)\;,
\elabel{orthotocart}
\ee
so that 
\[
\gamma_\mu^{(\rm in)}(\vc X_1)=-R_{\mu\nu}(2\vp_3-2\vp_1+\phi_1)\gamma_\nu^{(\rm o)}(\vc X_1)\;.
\]
Since $2\vp_3-2\vp_1+\phi_1=-2\pi+2\phi_2+\phi_1=-\pi+\phi_2-\phi_3$,
where we used Eq.\ (\ref{eq:angles}) and the fact that the sum of the
$\phi_l$ is $\pi$, one finally obtains
\be
\gamma_{\mu\nu\lambda}^{(\rm in)}=R_{\mu\alpha}(\phi_2-\phi_3)
R_{\nu\beta}(\phi_3-\phi_1)R_{\lambda\gamma}(\phi_1-\phi_2)
\gamma_{\alpha\beta\gamma}^{(\rm o)} \;,
\elabel{orthotoin}
\ee
where the result for the other two points were obtained from the one
at $\vc X_1$ by cyclic permutation of the indices. It should be noted
that in this equation, as in all the following, the angles $\phi_l$
are functions of the sidelengths $x_l$ (and the orientation), and not
independent geometrical quantities; hence, an equation like
(\ref{eq:orthotoin}) contains only the $x_l$ as independent variables.

\begin{figure}[!t]
  \parbox[t]{0.49\hsize}{%
    \resizebox{\hsize}{!}{\input fig3.tex}
    \caption{The circumcenter of a triangle is given by the
    intersection of the three midperpendiculars. According to Thales'
    theorem, the side $x_2$ subtends the angle $2\phi_2$ as seen from
    the circumcenter; from that, the angle between the lines
    connecting the points $\vc X_l$ with the circumcenter are easily
    obtained by noting that all three of them have equal length.
}%
    \label{fig:5}}
  \hfill
  \parbox[t]{0.49\hsize}{%
    \resizebox{\hsize}{!}{\input fig6.tex}
    \caption{The centers of the escribed circles are intersections
    between one interior angle bisector and two exterior angle
    bisectors. These three escribed circles are tangent to one side of
    the triangle and the extensions of the two other sides.
}%
    \label{fig:6}}
\end{figure}

Next we consider the projection onto the center of the
circumcircle. If $\zeta_1$ denotes the direction of the line
connecting $\vc X_1$ with this center, then the shear components with
respect to this center (denoted by the superscript `out') read 
\be
\gamma_\mu^{(\rm out)}=-R_{\mu\nu}(2\zeta_1)\gamma_\nu(\vc X_1)
=-R_{\mu\nu}(2\zeta_1-2\vp_1)\gamma_\nu^{(\rm o)}(\vc X_1)\;,
\ee
where in the second step we used Eq.\ (\ref{eq:orthotocart}). From Thales'
theorem (see Fig.\ts\ref{fig:5}) one finds that
$\zeta_1-\vp_3= \pm \pi/2 - \phi_3$, where the sign $\pm$ depends on
the orientation of the triangle.  Hence,
$\zeta_1-\vp_1= \pm \pi/2 - \phi_3 + \vp_3 - \vp_1= \mp\pi/2 + \phi_2
- \phi_3$, and we see that the sign ambiguity does not play a role
because of the rotation properties of the shear [cf.\ comment after
Eq.~\eqref{eq:transf}].  For the other two points, the corresponding
relations are obtained by cyclic permutations, so that 
\be
\gamma_{\mu\nu\lambda}^{(\rm out)}=R_{\mu\alpha}(2\phi_2-2\phi_3)
R_{\nu\beta}(2\phi_3-2\phi_1)R_{\lambda\gamma}(2\phi_1-2\phi_2)
\gamma_{\alpha\beta\gamma}^{(\rm o)} \;.
\ee
Finally, we consider the projection on the center of mass of the
triangle, or the centroid (denoted by the superscript `cen'), which
is the intersection of the three side-bisectors. Denoting by $\psi_3$
the angle between the side bisector of side $x_3$ and $\vc x_3$, we
obtain from Fig.\ts\ref{fig:4} that 
\[
\sin\psi_3={x_1\over h_3}\sin\phi_2={x_1 x_2\over h_3
x_3}\sin\phi_3\;,
\]
where we made use of the sine-theorem, and 
\[
h_3={1\over 2}\sqrt{2x_1^2+2x_2^2-x_3^2}
\]
is the length of the side-bisector of side $x_3$. From the same figure
we also obtain that
\[
\cos\psi_3={h_3^2+x_3^2/4-x_1^2\over h_3 x_3} ={x_2^2-x_1^2\over 2 h_3
x_3}\;,
\]
so that
\begin{align}
\cos 2\psi_3 = {} & {(x_2^2-x_1^2)^2-4 x_1^2x_2^2\sin^2\phi_3\over 4
h_3^2x_3^2}\;; &
\sin 2\psi_3 = {} & {(x_2^2-x_1^2)x_1x_2\sin\phi_3\over h_3^2x_3^2}\;.
\elabel{psitriag}
\end{align}
Analogous relations are obtained for the other two points by cyclic
permutations of the indices. The shear components projected toward the
centroid are then
\be
\gamma_\mu^{(\rm cen)}(\vc X_l)=R_{\mu\nu}(2\psi_l)\gamma^{(\rm s)}
=-R_{\mu\nu}(2\psi_l)\gamma^{(\rm o)}\;,
\ee
where the components of the rotation matrix can be directly obtained
from Eqs.\ (\ref{eq:psitriag}). For completeness, we give the 3PCF for the
centroid,
\be
\gamma_{\mu\nu\lambda}^{(\rm cen)}=-R_{\mu\alpha}(2\psi_1)
R_{\nu\beta}(2\psi_2)R_{\lambda\gamma}(2\psi_3)
\gamma_{\alpha\beta\gamma}^{(\rm o)} \;.
\ee

\section{Invariant components of the shear three-point correlation
function} 
The normal way to choose `good' components of a multi-index quantity
like $\gamma_{\mu\nu\lambda}$ is to look for its behavior under
coordinate transformations. Recall the situation for the shear
two-point correlation function: the correlator $\ave{\gamma_\mu
\gamma_\nu}(\vc x)$ of the Cartesian components of the shear contains
two terms, one which is independent on the phase $\vp$ of $\vc x$, the
other behaving as $\cos(4\vp)$. The coefficients of these two terms
are $\xi_+$ and $\xi_-$, respectively, which are the invariants, and
therefore the natural choice for the components of the two-point
correlation function of the shear.

Similarly, we can consider the behavior of the 3PCF under
rotations. Given that no linear scalar can be built from the Cartesian
components of the 3PCF alone we cannot expect to find a combination of
components which is invariant under general rotations. However, as we
shall see, there are linear combinations of the components of the 3PCF
which have a simple behavior under rotations, namely multiplication by
a phase factor. Owing to this simple transformation behavior, we shall
term them `natural' components.

Let now $\gamma_{\mu\nu\lambda}$ be the components of the 3PCF of
the shear measured with respect to one particular choice of
direction(s); choosing different projection directions, which differ
from the old ones by $\zeta_i$, the 3PCF becomes
\be
\gamma'_{\mu\nu\lambda}\equiv\ave{\gamma'_\mu(\vc X_1)\gamma'_\nu(\vc
X_2)\gamma'_\lambda(\vc X_3)}= 
R_{\mu\alpha}(2\zeta_1)\,R_{\nu\beta}(2\zeta_2)\,R_{\lambda\gamma}(2\zeta_3) 
\gamma_{\alpha\beta\gamma}\;;
\elabel{rotation}
\ee
note that this expression differs from Eq.\ (\ref{eq:transfo}) by a minus
sign since in Eq.\ (\ref{eq:transfo}) the transition from Cartesian to
tangential and cross components was included as well. To investigate
more closely the action of the rotation operator in
Eq.\ (\ref{eq:rotation}), we transform this equation into one which appears
more familiar: We define the eight-component quantity
\[
\Gamma:=(\gamma_{{\rm ttt}},\gamma_{{\rm tt}\times},
\gamma_{\rm t\times t},\gamma_{\rm t\times \times},
\gamma_{{\rm \times tt}},\gamma_{{\rm \times t}\times},
\gamma_{\rm \times\times t},\gamma_{\rm \times\times \times})
\]
and analogously $\Gamma'$ for the transformed components; then, the
rotation described in Eq.\ (\ref{eq:rotation}) can be written as $\Gamma'=R
\Gamma$, where $R$ is a $8\times 8$ matrix. The components of $R$ are
triple products of trigonometric functions. Since it describes a
rotation, $R$ is unitary and 
one expects that the eigenvalues of $R$ have an absolute
value of unity. They can in fact be obtained as
\begin{align}
\lambda^{(0)}_{1,2} = {} &\exp\rund{\pm 2{\rm
    i}[\zeta_1+\zeta_2+\zeta_3]}\; ; & 
\lambda^{(1)}_{1,2} = {} &\exp\rund{\pm 2{\rm
i}[-\zeta_1+\zeta_2+\zeta_3]} \; ; \nonumber \\
\lambda^{(2)}_{1,2} = {} &\exp\rund{\pm 2{\rm
    i}[\zeta_1-\zeta_2+\zeta_3]} \; ; & 
\lambda^{(3)}_{1,2} = {} & \exp\rund{\pm 2{\rm
    i}[\zeta_1+\zeta_2-\zeta_3]} \; .
\elabel{EV}
\end{align}
Note that this result is not very surprising: either the eigenvalues
are $\pm 1$, or they have to appear in the above form, i.e., they have
to occur as pairs of complex conjugate numbers with absolute value of
unity. The dependence of the eigenvalues on the rotation angles is in
fact a natural one.

\subsection{The natural components of the 3PCF}
We see that under the general rotation described by
Eq.\ (\ref{eq:rotation}), the eigenvalue $+1$ does not occur; in other
words, there is no (linear) combination of the components of the 3PCF
that is invariant under the transformation
(\ref{eq:rotation}). However, for some special rotations, the
eigenvalue $+1$ occurs. As a first example, we consider the case
$\zeta_1+\zeta_2+\zeta_3=0$; as shown in Sect.\ts 3.2, this case
actually is encountered for transformations of the shear components
between the orthocenter, the incenter and the circumcenter. Then, the
two eigenvalues $\lambda^{(0)}_{1,2}$ are $+1$, and the corresponding
eigenvectors can be found to be
\begin{align}
E^{(0)}_1 = {} & (1,0,0,-1,0,-1,-1,0) \; ; &
E^{(0)}_2 = {} & (0, 1, 1, 0, 1, 0, 0, -1) \; .
\end{align}
Hence, under rotations of this kind, we expect that the two
combinations
\begin{align}
\Gamma^{(0)}_1 = {} & \gamma_{{\rm ttt}}-\gamma_{\rm t\times
  \times}-\gamma_{{\rm \times t}\times}-\gamma_{{\times}{\times}\rm t}
  \; ; &
\Gamma^{(0)}_2 = {} & \gamma_{{\rm tt}\times}+\gamma_{\rm t\times t}
  + \gamma_{{\rm \times tt}}-\gamma_{\rm \times\times\times}
\end{align}
of the components of the 3PCF are both invariant. It will turn out to
be very useful to combine these two invariants into a single complex
quantity, $\Gamma^{(0)}=\Gamma^{(0)}_1+{\rm i} \Gamma^{(0)}_2$. This quantity,
however, can also be written in a different form, namely
\be
\Gamma^{(0)}=\ave{\gamma(\vc X_1)\gamma(\vc X_2)\gamma(\vc X_3)}\;,
\elabel{inva00}
\ee
where we again consider $\gamma=\gamma_{\rm t}+{\rm i}\gamma_\times$
as a complex quantity. Written in this form, it is obvious that a
rotation with $\zeta_1+\zeta_2+\zeta_3=0$ keeps $\Gamma^{(0)}$ invariant.

Next we consider the case that $-\zeta_1+\zeta_2+\zeta_3=0$. Such a
rotation occurs if one transforms the shear components from the
incenter (or the out- or orthocenter) to the center of one of the
escribed circles (see Fig.\ts \ref{fig:6} for an explanation); in this case,
$\lambda^{(1)}_{1,2}=1$, and from the corresponding eigenvectors one
can again construct two combinations $\Gamma^{(1)}_1$, $\Gamma^{(1)}_2$, of the
components of the 3PCF that stay invariant under such a rotation, and
they can as well be combined into a single complex quantity. The same
procedure can then be repeated for rotations with
$\zeta_1-\zeta_2+\zeta_3=0$ and those with
$\zeta_1+\zeta_2-\zeta_3=0$; these correspond, e.g., to the
transformation from one of the aforementioned centers to the centers
of the other two escribed circles. The corresponding invariants read:
\begin{align}
\Gamma^{(1)} = {} &\gamma_{{\rm ttt}}-\gamma_{\rm t\times
  \times}+\gamma_{{\rm \times t}\times}+\gamma_{{\times}{\times}\rm t}
+{\rm i}\eck{\gamma_{{\rm tt}\times}+\gamma_{\rm t\times t}
-\gamma_{{\rm \times tt}}+\gamma_{\rm \times\times\times}}
=\ave{\gamma^*(\vc X_1)\gamma(\vc X_2)\gamma(\vc X_3)}\; ; \nonumber \\
\Gamma^{(2)} = {} &\gamma_{{\rm ttt}}+\gamma_{\rm t\times \times}-\gamma_{{\rm
\times t}\times}+\gamma_{{\times}{\times}\rm t}
+{\rm i}\eck{\gamma_{{\rm tt}\times}-\gamma_{\rm t\times t}
+\gamma_{{\rm \times tt}}+\gamma_{\rm \times\times\times}}
=\ave{\gamma(\vc X_1)\gamma^*(\vc X_2)\gamma(\vc X_3)}\;; 
\elabel{invariants}\\
\Gamma^{(3)} = {} &\gamma_{{\rm ttt}}+\gamma_{\rm t\times \times}+\gamma_{{\rm
\times t}\times}-\gamma_{{\times}{\times}\rm t}
+{\rm i}\eck{-\gamma_{{\rm tt}\times}+\gamma_{\rm t\times t}
+\gamma_{{\rm \times tt}}+\gamma_{\rm \times\times\times}}
=\ave{\gamma(\vc X_1)\gamma(\vc X_2)\gamma^*(\vc X_3)}\;. \nonumber
\end{align}
Now, each of these $\Gamma^{(\alpha)}$ is invariant only under special
rotations, as the derivation above has shown. But the remarkable
finding here is that, under a general rotation, the different
$\Gamma^{(\alpha)}$ do not mix, but they are just multiplied by a
phase factor; indeed, since the rotation (\ref{eq:rotation}) implies
that $\gamma(\vc X_1)\to\gamma'(\vc X_1)=\gamma(\vc X_1)\,{\rm
e}^{-2{\rm i}\zeta_1}$, and similarly for the other two points, we see
from the final expression in Eqs.\ (\ref{eq:inva00}) and
(\ref{eq:invariants}) that the
transformed invariants read
\be
(\Gamma^{(0)})'=\exp\rund{- 2{\rm i}[\zeta_1+\zeta_2+\zeta_3]}\Gamma^{(0)}
=\lambda^{(0)}_2\,\Gamma^{(0)} \;,\; {\rm and}\;
(\Gamma^{(\alpha)})'=\lambda^{(\alpha)}_2\,\Gamma^{(\alpha)}\;,
\elabel{invtrans}
\ee
where the $\lambda^{(\alpha)}_2$ have been defined in Eq.\ (\ref{eq:EV}).
The transformation (\ref{eq:invtrans}) is indeed remarkable, though in
hindsight not all that surprising. The fact that the
$\Gamma^{(\alpha)}$ only transform amongst themselves justifies that
they are called {\it natural components} of the three-point
correlation function of the shear. {\it Note in particular that
Eq.\ (\ref{eq:invtrans}) implies that the four absolute values
$\abs{\Gamma^{(\alpha)}}$ are independent under general rotations;
they are therefore the (non-linear) invariants of the shear 3PCF.}

Of course, if desired, the original components of the shear 3PCF can
be obtained from the natural components; the inverse of
Eq.\ (\ref{eq:invariants}) reads:
\begin{align}
\gamma_{{\rm ttt}} = {} &\rund{\Gamma^{(1)}_1+\Gamma^{(2)}_1
+\Gamma^{(3)}_1+\Gamma^{(0)}_1}/4\; ; &
\gamma_{\rm t\times \times} = {} & \rund{-\Gamma^{(1)}_1+\Gamma^{(2)}_1
+\Gamma^{(3)}_1-\Gamma^{(0)}_1}/4\; ;
\nonumber \\
\gamma_{{\rm \times t}\times} = {} &\rund{\Gamma^{(1)}_1-\Gamma^{(2)}_1
+\Gamma^{(3)}_1-\Gamma^{(0)}_1}/4\; ; &
\gamma_{{\times}{\times}\rm t} = {} & \rund{\Gamma^{(1)}_1+\Gamma^{(2)}_1
-\Gamma^{(3)}_1-\Gamma^{(0)}_1}/4\; ; \nonumber \\
\gamma_{{\rm tt}\times} = {} &\rund{\Gamma^{(1)}_2+\Gamma^{(2)}_2
-\Gamma^{(3)}_2+\Gamma^{(0)}_2}/4\; ; &
\gamma_{\rm t\times t} = {} &\rund{\Gamma^{(1)}_2-\Gamma^{(2)}_2
+\Gamma^{(3)}_2+\Gamma^{(0)}_2}/4\; ; \\
\gamma_{{\rm \times tt}} = {} &\rund{-\Gamma^{(1)}_2+\Gamma^{(2)}_2
+\Gamma^{(3)}_2+\Gamma^{(0)}_2}/4\; ; &
\gamma_{\rm \times\times\times} = {} & \rund{\Gamma^{(1)}_2+\Gamma^{(2)}_2
+\Gamma^{(3)}_2-\Gamma^{(0)}_2}/4\; .\nonumber
\end{align}

\subsection{Parity transformation}
A parity transformation is obtained by mirroring a given set of three
points with respect to any straight line. Such a transformation has
the following effects: if in the original triangle the three points
$\vc X_1$, $\vc X_2$, $\vc X_3$ are ordered counterclockwise with
respect to, say, the center of the incircle, in the transformed
triangle they will be ordered clockwise.  As a result, under a parity
transformation we need to replace every angle with its opposite.  

In order to understand the behavior of the 3PCF under parity, let us
suppose that the triangle is flipped along a line perpendicular to the
side $\vc x_3$ (since the 3PCF is invariant upon rotation, the
direction used for the flipping is irrelevant).  We first observe that
this mirror symmetry is equivalent to interchanging the points $\vc
X_1$ and $\vc X_2$, or the sides $x_1$ and $x_2$.  Second, such a
flipping of orientation keeps the tangential component of the shear
invariant (if the direction relative to which the shear components are
measured are as well subject to the mirror transformation -- like it
happens when shear components are defined relative to one of the
centers of the triangle), but changes the sign of the cross-component;
hence, this transformation implies ${\rm P}\gamma
=\gamma^*$. Together, these two effects therefore imply that under a
parity transformation, 
\be 
{\rm P}\eck{\gamma_{\mu\nu\lambda}(x_1,x_2,x_3)} =\Pi
\gamma_{\nu\mu\lambda}(x_2,x_1,x_3)\; ,\quad {\rm where}\quad
\Pi=(-1)^{\nu+\mu+\lambda+1}\;, \elabel{paritys} 
\ee 
the parity, is $+1$ if all of the indices of $\gamma$ are ${\rm t}$'s,
or two $\times$'s occur, otherwise it is negative.  Components of the
3PCF with $\Pi=+1$ are called even, those of negative parity odd
components.  The action of the parity transformation is most easily
expressed in terms of the natural components of the 3PCF; as can be
seen from Eq.\ (\ref{eq:invariants}), the real part of each of the
$\Gamma^{(\alpha)}$ is composed of even components, the imaginary part
of odd components. Therefore,
\begin{align}
{\rm P}\eck{\Gamma^{(0)}(x_1,x_2,x_3)} = {} &
\rund{\Gamma^{(0)}}^*(x_1,x_3,x_2)=
\rund{\Gamma^{(0)}}^*(x_2,x_1,x_3)=
\rund{\Gamma^{(0)}}^*(x_3,x_2,x_1) \;,\nonumber \\
{\rm P}\eck{\Gamma^{(1)}(x_1,x_2,x_3)} = {} &
\rund{\Gamma^{(1)}}^*(x_1,x_3,x_2)=
\rund{\Gamma^{(2)}}^*(x_2,x_1,x_3)=
\rund{\Gamma^{(3)}}^*(x_3,x_2,x_1) \;,\nonumber \\
{\rm P}\eck{\Gamma^{(2)}(x_1,x_2,x_3)} = {} &
\rund{\Gamma^{(1)}}^*(x_2,x_1,x_3)=
\rund{\Gamma^{(2)}}^*(x_3,x_2,x_1)=
\rund{\Gamma^{(3)}}^*(x_1,x_3,x_2) \;, \\
{\rm P}\eck{\Gamma^{(3)}(x_1,x_2,x_3)} = {} &
\rund{\Gamma^{(1)}}^*(x_3,x_2,x_1)=
\rund{\Gamma^{(2)}}^*(x_1,x_3,x_2)=
\rund{\Gamma^{(3)}}^*(x_2,x_1,x_3) \;.\nonumber 
\end{align}
Each of these parity transformations contains a permutation of the
arguments with negative signature. One can also consider cyclic
permutations of the arguments; since
\be
\gamma_{\mu\nu\lambda}(x_1,x_2,x_3)
=\gamma_{\nu\lambda\mu}(x_2,x_3,x_1)
=\gamma_{\lambda\mu\nu}(x_3,x_1,x_2)
\;,
\ee
one obtains for the natural components
\begin{align}
\Gamma^{(0)}(x_1,x_2,x_3) = {}
&\Gamma^{(0)}(x_2,x_3,x_1)=\Gamma^{(0)}(x_3,x_1,x_2) 
\;,\nonumber \\
\Gamma^{(1)}(x_1,x_2,x_3) = {}
&\Gamma^{(3)}(x_2,x_3,x_1)=\Gamma^{(2)}(x_3,x_1,x_2) \; ,\nonumber \\
\Gamma^{(2)}(x_1,x_2,x_3) = {}
&\Gamma^{(1)}(x_2,x_3,x_1)=\Gamma^{(3)}(x_3,x_1,x_2) \; ,
\elabel{permuts}
\\
\Gamma^{(3)}(x_1,x_2,x_3) = {}
&\Gamma^{(2)}(x_2,x_3,x_1)=\Gamma^{(1)}(x_3,x_1,x_2)
\;.\nonumber 
\end{align}
Hence, $\Gamma^{(0)}$ is invariant under cyclic permutations of the
arguments, whereas the other three natural components of the shear
3PCF transform into each other under such permutations. The relations
(\ref{eq:permuts}) imply that only one of the three functions
$\Gamma^{(k)}(x_1,x_2,x_3)$, $1\le k\le 3$, is independent, the other
two can be obtained by cyclic permutations of the arguments.

The parity transformations have an immediate consequence for triangles
where two sides are equal, say $x_1=x_2$; namely, from
Eq.\ (\ref{eq:paritys}) one finds that
\begin{align}
\gamma_{\rm
tt\times}(x_1,x_1,x_3)= {} & 0=\gamma_{\times\times\times}(x_1,x_1,x_3)\;
; & 
\gamma_{\rm t\times t}(x_1,x_1,x_3)= {} & -\gamma_{\rm \times
tt}(x_1,x_1,x_3)\;.
\end{align}
This implies that for $x_1=x_2$, $\Gamma^{(0)}$ and $\Gamma^{(3)}$
have no imaginary part, and those of $\Gamma^{(1)}$ and $\Gamma^{(2)}$
have equal magnitude but opposite sign.  Furthermore, for equilateral
triangles, all odd components of the shear 3PCF vanish, in which case
the natural components become purely real.

\subsection{Two-point correlation function revisited}
The foregoing formalism can of course also be applied to the two-point
correlation function. In that case, the eigenvalues are
$\lambda^{(0,1)}_{1,2}= {\rm e}^{\pm 2{\rm i}(\zeta_1\pm\zeta_2)}$,
and the invariant combinations are $\Gamma^{(0)}=\gamma_{\rm
tt}-\gamma_{\times\times} +{\rm i}\rund{\gamma_{\rm t\times}
+\gamma_{\rm \times t}}$ and 
$\Gamma^{(1)}=\gamma_{\rm
tt}+\gamma_{\times\times} +{\rm i}\rund{\gamma_{\rm t\times}
-\gamma_{\rm \times t}}$. The two-point correlator is, however,
special in the sense that after a parity transformation, the two
points can be brought back into the old positions with a
rotation. This then implies that the imaginary components of the
$\Gamma^{(\alpha)}$ vanish if they are measured in the only reference
frame that makes sense for two points -- namely, the line connecting
them; and so $\Gamma^{(0)}=\xi_-$, $\Gamma^{(1)}=\xi_+$.
Note that if different projection directions are taken, the
imaginary parts of the $\Gamma^{(\alpha)}$ are not zero.

\subsection{Generalization}
The discussion on obtaining the natural components of the shear 3PCF
immediately suggests how to generalize it to the natural components of
higher-order correlation functions. Here, we shall give the results
for the four-point function (also see the recent work by Takada \& Jain
2002a for a detailed consideration of the kurtosis of the cosmic shear
field); further generalizations are straightforward to obtain:

If $\gamma_{\mu\nu\lambda\kappa}$ is the four-point correlation
function of the shear (for notational simplicity, we skip the
arguments of this function), and we consider a general rotation of the
directions relative to which tangential and cross components are
defined, the transformation of the shear four-point correlation
function reads
\be
\gamma'_{\mu\nu\lambda\kappa}\equiv\ave{\gamma'_\mu(\vc X_1)\gamma'_\nu(\vc
X_2)\gamma'_\lambda(\vc X_3)\gamma'_\kappa(\vc X_4)}= 
R_{\mu\alpha}(2\zeta_1)\,R_{\nu\beta}(2\zeta_2)\,R_{\lambda\gamma}(2\zeta_3)\,
R_{\kappa\delta}(2\zeta_4) 
\gamma_{\alpha\beta\gamma\delta}\;.
\elabel{4-rotation}
\ee
In analogy to the treatment for the 3PCF, there are now eight 
complex conjugate pairs of eigenvalues, which are
\begin{align}
\lambda^{(0)}_{1,2} = {} &\exp\rund{\pm 2{\rm
    i}[\zeta_1+\zeta_2+\zeta_3+\zeta_4]}\;; &
\lambda^{(1)}_{1,2} = {} & \exp\rund{\pm 2{\rm
i}[\zeta_1+\zeta_2-\zeta_3-\zeta_4]}\;;\nonumber \\
\lambda^{(2)}_{1,2} = {} &\exp\rund{\pm 2{\rm i}
[\zeta_1-\zeta_2+\zeta_3-\zeta_4]}\;; &
\lambda^{(3)}_{1,2} = {} & \exp\rund{\pm 2{\rm i}
[-\zeta_1+\zeta_2+\zeta_3-\zeta_4]}\;;\nonumber \\
\lambda^{(4)}_{1,2} = {} & \exp\rund{\pm 2{\rm
i}[\zeta_1+\zeta_2+\zeta_3-\zeta_4]}\;; &
\lambda^{(5)}_{1,2} = {} & \exp\rund{\pm 2{\rm i}
[\zeta_1+\zeta_2-\zeta_3+\zeta_4]}\;; \elabel{4-EV}
\\
\lambda^{(6)}_{1,2} = {} &\exp\rund{\pm 2{\rm i}
[\zeta_1-\zeta_2+\zeta_3+\zeta_4]}\;; &
\lambda^{(7)}_{1,2} = {} & \exp\rund{\pm 2{\rm i}
[-\zeta_1+\zeta_2+\zeta_3+\zeta_4]}\;.\nonumber 
\end{align}
The corresponding natural components are then
\begin{align}
\Gamma^{(0)} = {} &\ave{\gamma(\vc X_1)\gamma(\vc X_2)\gamma(\vc
X_3)\gamma(\vc X_4)}\;;&
\Gamma^{(1)} = {} & \ave{\gamma(\vc X_1)\gamma(\vc X_2)\gamma^*(\vc
X_3)\gamma^*(\vc X_4)}\;;\nonumber \\ 
\Gamma^{(2)} = {} &\ave{\gamma(\vc X_1)\gamma^*(\vc X_2)\gamma(\vc
X_3)\gamma^*(\vc X_4)}\;; &
\Gamma^{(3)} = {} & \ave{\gamma^*(\vc X_1)\gamma(\vc X_2)\gamma(\vc
X_3)\gamma^*(\vc X_4)}\;;\nonumber \\ 
\Gamma^{(4)} = {} &\ave{\gamma(\vc X_1)\gamma(\vc X_2)\gamma(\vc
X_3)\gamma^*(\vc X_4)}\;; &
\Gamma^{(5)} = {} & \ave{\gamma(\vc X_1)\gamma(\vc X_2)\gamma^*(\vc
X_3)\gamma(\vc X_4)}\;; \\ 
\Gamma^{(6)} = {} &\ave{\gamma(\vc X_1)\gamma^*(\vc X_2)\gamma(\vc
X_3)\gamma(\vc X_4)}\;; &
\Gamma^{(7)} = {} & \ave{\gamma^*(\vc X_1)\gamma(\vc X_2)\gamma(\vc
X_3)\gamma(\vc X_4)}\;. \nonumber
\end{align}
For reference, we shall write down the first of these explicitly,
\begin{align}
\Gamma^{(0)} = {} &\gamma_{\rm tttt}+\gamma_{\times\times\times\times}
-\gamma_{\rm tt\times\times}-\gamma_{\rm t\times t\times}
-\gamma_{\rm t \times\times t} -\gamma_{\rm \times t t \times}
-\gamma_{\rm \times t \times t} -\gamma_{\rm \times \times t t}
\nonumber \\
& {} + {\rm i}\eck{\gamma_{\rm ttt\times}+\gamma_{\rm tt\times t}
+\gamma_{\rm t\times tt}+\gamma_{\rm \times ttt}
-\gamma_{\rm t \times\times \times} -\gamma_{\rm \times t \times \times}
-\gamma_{\rm \times  \times t \times} -\gamma_{\rm \times \times
\times t}}
\end{align}
These natural components of the shear four-point correlation function
will transform under the rotation (\ref{eq:4-rotation}) like
\be
\rund{\Gamma^{(\alpha)}}'=\lambda_2^{(\alpha)}\,\Gamma^{(\alpha)}\;.
\ee
These relations make it obvious how generalizations to higher-order
correlations can be obtained.

\subsection{The estimator of Bernardeau et al.\ (2002)}
In their paper, Bernardeau et al.\ (2002b) considered, for the first
time, a specific shear 3PCF, which they then applied successfully to
observational cosmic shear data (Bernardeau et al.\ 2002a). Here, we
shall write the Bernardeau et al.\ (2002b) estimator in our
notation. They considered one side of a triangle (say, $\vc x_3$) as
the reference direction, and project the shear at all three points
along this direction. Hence, their projected components (here written
as $\gamma^{(3)}_{\rm t,\times}$) read
\be
\gamma^{(3)}_\mu(\vc X_l)=-R_{\mu\nu}(2\vp_3)\gamma_\nu(\vc X_l)
=R_{\mu\nu}(2\vp_3-2\vp_l)\gamma_\nu^{(\rm s)}(\vc X_l)\;,
\ee
and, of course, $\gamma^{(3)}(\vc X_3)=\gamma^{(\rm s)}(\vc X_3)$.
The 3PCF defined by Bernardeau et al.\ (2002b) then becomes
\be
\Gamma_\mu^{(\rm B)}=\ave{\eck{\gamma_{\rm t}^{(3)}(\vc
X_1)\gamma_{\rm t}^{(3)}(\vc X_2)+\gamma_{\times}^{(3)}(\vc
X_1)\gamma_{\times}^{(3)}(\vc X_2)}
\gamma_\mu^{(3)}(\vc X_3)}\;.
\ee
Since
\begin{align}
& \gamma_{\rm t}^{(3)}(\vc X_1) \gamma_{\rm t}^{(3)}(\vc X_2) +
\gamma_{\times}^{(3)}(\vc X_1)\gamma_{\times}^{(3)}(\vc X_2) \nonumber \\
&\qquad {} = \eck{R_{1\nu}(2\vp_3-2\vp_1)R_{1\lambda}(2\vp_3-2\vp_2)
+R_{2\nu}(2\vp_3-2\vp_1)R_{2\lambda}(2\vp_3-2\vp_2)}\gamma_\nu^{(\rm
s)}(\vc X_1)\gamma_\lambda^{(\rm s)}(\vc X_2)\nonumber \\
&\qquad {} = R_{\nu\lambda}(2\vp_1-2\vp_2)\gamma_\nu^{(\rm
s)}(\vc X_1)\gamma_\lambda^{(\rm s)}(\vc X_2)\;,\nonumber
\end{align}
one obtains
\be
\Gamma_\mu^{(\rm B)}=R_{\nu\lambda}(2\phi_3)
\gamma^{(\rm s)}_{\nu\lambda\mu}\;,
\ee
provided the triangle formed by the points $\vc X_l$ has the
orientation defined at the beginning of Sect.\ts 3.

\section{Discussion and outlook}
The natural components of the shear 3PCF provide a generalization of
the corresponding natural components $\xi_\pm$ of the two-point
correlation function. Given that upcoming cosmic shear surveys will be
substantially larger than the current ones, it is obvious that the
3PCF will be measurable in the future with a similar accuracy as the
two-point correlation function in current surveys; in fact, a first
significant measurement of the 3PCF has been reported in Bernardeau et
al.\ (2002a). Therefore, it is worth to explore the dependence of the
3PCF on various parameters, as has been done for the two-point
function. In particular, the detailed study of the interrelations
between $\xi_\pm$ and the underlying power spectrum, as presented in
Crittenden et al.\ (2002) and Schneider et al.\ (2002a), shall be
generalized to the 3PCF. Of course, this will be substantially more
difficult from a technical point of view, and will be deferred to
future work. Nevertheless, we can outline a few aspects of
what can be expected from such work.

In close analogy to the two-point correlation function, one can expect
that the natural components can be more easily calculated
from the bispectrum of the underlying mass distribution than the
individual components of the 3PCF. As is the case for $\xi_\pm$, one
can expect that $\Gamma^{(0)}$ probes the bispectrum in a different
way than the $\Gamma^{(k)}$, $1\le k\le 3$. In particular, the
different natural components will have a different dependence on
cosmological parameters.

The natural components of the 3PCF are not independent of each other;
provided that the shear indeed is due to a surface mass density field,
there should be integral relations which interrelate them. Again one
should note the analogy with the two-point correlation function, where
$\xi_+$ can be obtained as an integral over $\xi_-$ and vice versa
(Crittenden et al.\ 2002; Schneider et al.\ 2002a). The interrelations
between the components of the 3PCF provide a redundancy which can be
profitably combined to reduce the noise in real measurements.

If the shear is not solely due to a surface mass distribution, the
shear field may contain a B-mode contribution (e.g., from intrinsic
alignments of the galaxies from which the shear is measured). In the
case of the two-point correlation function, the presence of a B-mode
can be probed from integral relations between the two correlation
functions $\xi_\pm$ (Crittenden et al.\ 2002; Schneider et al.\ 2002a);
it is expected that similarly in the case of the 3PCF, the integral
relations between the natural components will be modified in the
presence of a B-mode.

All linear three-point statistics of the shear can be expressed in
terms of the shear 3PCF. This is obvious from the fact that the
bispectrum of the surface mass density can be expressed in terms of the
3PCF; on the other hand, all linear three-point statistics are
linearly related to the bispectrum, and can therefore be expressed
directly in terms of the 3PCF. In particular, the third-order aperture
mass statistics (Schneider et al.\ 1998; van Waerbeke et al.\ 1999)
can be expressed as an integral over the 3PCF. It remains to be seen
whether the third-order aperture mass statistics is as useful for a
separation of the shear field into E- and B-modes as it is the case
for the two-point statistics (Crittenden et al.\ 2002; Schneider et
al.\ 2002a; for applications in cosmic shear surveys, see e.g.\ Pen et
al.\ 2002; Hoekstra et al.\ 2002).

Concerning practical measurements of the 3PCF, the procedure is
straightforward: from a given catalog of galaxy images with position
vectors $\vc \theta_i$ and measured ellipticities $\eps_i$, triplets
are selected. For each such triplet, the sides of the corresponding
triangle can be calculated, and for practical reasons the largest side
be called $x_3$. The three points are then labeled $\vc X_l$, $1\le
l\le 3$ in the (unique) way such that the orientation of the points is
as described at the beginning of Sect.\ts 3, and the longest side
connects $\vc X_1$ and $\vc X_2$. Since the connecting vectors $\vc
x_l$ need to be calculated anyway, it is simplest to project the
ellipticities along the sides of the triangle, or, with opposite sign,
towards the orthocenter, without having to calculate any trigonometric
function. The eight ($=2\times 2\times 2$) triple products of the
projected ellipticities are calculated and summed up in eight 
three-dimensional bins. Those are conveniently labeled by a scale
$x_3$, and two shape parameters, $q_1=x_1/x_3$, $q_2=x_2/x_3$, so that
the grid of bins runs as $0\le q_{1,2}\le 1$, $0<x_3<\infty$. After
summing up over all triplets of points, for each of the eight
components the sums in the bins are divided by the number of triplets
contributing, which then yields an estimate of the 3PCF. Those can
then be combined into the natural components for further analysis. 

For a quantitative analysis of the three-point statistics of cosmic
shear in relation to predictions of cosmological models, one can
either use integrated properties of the shear 3PCF -- such as the
aperture mass or the integral of $\Gamma^{(\rm B)}$ over an elliptical
region as done in Bernardeau et al.\ (2002b), or consider the (more
noisy, but much more numerous) estimates of the natural components of
the 3PCF directly. The first method yields one-dimensional functions
of the three-point shear statistics depending on a scale parameter,
and are thus very convenient for graphical displays. In contrast,
using the full shear 3PCF employs multi-dimensional data which is
difficult to display, but contains {\it all} the information about
third-order statistics from the data. Hence, even though the
signal-to-noise in each bin of the 3PCF can be small, its overall
information content cannot be smaller than that of any of the
integrated quantities, and should therefore be employed in extracting
cosmological information from the measurements. To obtain a reliable
figure-of-merit for the comparison of the observational results with
model predictions, one needs to know the covariance matrix of the
shear 3PCF -- this, however, is a 64-components quantity depending on
6 arguments, and will be very difficult to obtain
analytically (see Schneider et al.\ 2002b for the difficulties of
obtaining the covariance of the two-point correlation functions). 
Corresponding covariances of integrated quantities may
be slightly easier obtainable. On the other hand, it is quite likely
that figures-of-merit will have to rely heavily on future ray-tracing
simulations through N-body-generated cosmological mass distributions,
as in, e.g., van Waerbeke et al.\ (1999), Jain et al.\ (2000) and
Bernardeau et al.\ (2002a). 

After this paper was finished, two recent preprints were posted on the
Web which discussed very similar issues (Zaldarriaga \& Scoccimarro
2002; Takada \& Jain 2002b). In the latter one, the
components $\gamma_{\mu\nu\lambda}$ of the correlation function were
calculated from ray-tracing simulations; in particular they
verified that all of the eight components of the 3PCF are non-zero in
general and thus contain cosmological information. 

\begin{acknowledgements}
We are grateful to Matthias Bartelmann for very useful comments on
this manuscript.
This work was supported by the TMR Network ``Gravitational Lensing:
New Constraints on Cosmology and the Distribution of Dark Matter'' of
the EC under contract No. ERBFMRX-CT97-0172, by the German Ministry
for Science and Education (BMBF) through the DLR under the project 50
OR 0106, and the Deutsche Forschungsgemeinschaft under the project
SCHN 342/3--1.

\end{acknowledgements}

\end{document}

%% file: fig1.tex
\SpecialCoor
\def\Ax{1 }\def\Ay{0 }%
\def\Bx{9 }\def\By{2 }%
\def\Cx{2 }\def\Cy{6 }%
\psset{unit=0.8cm}%
\begin{pspicture}(0,-0.2)(11.5,7.2)
  \pnode(! \Ax \Ay){X_1}
  \pnode(! \Bx \By){X_2}
  \pnode(! \Cx \Cy){X_3}
  \pnode(! \Bx \Ax sub \By \Ay sub){x_3}
  \pnode(! \Cx \Bx sub \Cy \By sub){x_1}
  \pnode(! \Ax \Cx sub \Ay \Cy sub){x_2}
  \ncline[linewidth=1.2pt]{*->}{X_1}{X_2}
  \Bput{$\vec{x}_3$}
  \ncline[linewidth=1.2pt]{*->}{X_2}{X_3}
  \Bput{$\vec{x}_1$}
  \ncline[linewidth=1.2pt]{*->}{X_3}{X_1}
  \Bput{$\vec{x}_2$}
  \uput[180](X_1){$\vec{X}_1$}
  \uput[-60](X_2){$\vec{X}_2$}
  \uput[90](X_3){$\vec{X}_3$}
  \pcline[linestyle=dashed]{-}(X_1)([nodesep=2]X_1)
  \psarc{->}(X_1){1.5}{0}{(x_3)}
  \uput{1.7}[! \By \Ay sub \Bx \Ax sub atan 2 div](X_1){$\varphi_3$}
  \pcline[linestyle=dashed]{-}(X_2)([nodesep=2]X_2)
  \psarc{->}(X_2){1.5}{0}{(x_1)}
  \uput{1.7}[! \Cy \By sub \Cx \Bx sub atan 2 div](X_2){$\varphi_1$}
  \pcline[linestyle=dashed]{-}(X_3)([nodesep=2]X_3)
  \psarc{->}(X_3){1.5}{0}{(x_2)}
  \uput{1.7}[! \Ay \Cy sub \Ax \Cx sub atan 2 div](X_3){$\varphi_2$}
  \psarc(X_1){1}{(x_3)}{! \Cy \Ay sub \Cx \Ax sub atan}
  \uput{1.2}[! \By \Ay sub \Bx \Ax sub atan 
  \Cy \Ay sub \Cx \Ax sub atan add 2 div](X_1){$\phi_1$}
  \psarc(X_2){1}{(x_1)}{! \Ay \By sub \Ax \Bx sub atan}
  \uput{1.2}[! \Cy \By sub \Cx \Bx sub atan 
  \Ay \By sub \Ax \Bx sub atan add 2 div](X_2){$\phi_2$}
  \psarc(X_3){1}{(x_2)}{! \By \Cy sub \Bx \Cx sub atan}
  \uput{1.2}[! \Ay \Cy sub \Ax \Cx sub atan 
  \By \Cy sub \Bx \Cx sub atan add 2 div](X_3){$\phi_3$}
\end{pspicture}%
%

%% file: fig4.tex
\SpecialCoor
\def\Ax{1 }\def\Ay{0 }%
\def\Bx{9 }\def\By{2 }%
\def\Cx{2 }\def\Cy{6 }%
\def\a{\Bx \Cx sub dup mul \By \Cy sub dup mul add sqrt }%
\def\b{\Cx \Ax sub dup mul \Cy \Ay sub dup mul add sqrt }%
\def\c{\Ax \Bx sub dup mul \Ay \By sub dup mul add sqrt }%
\def\A{\Cy \Ay sub \Cx \Ax sub atan \By \Ay sub \Bx \Ax sub atan sub }%
\def\B{\Ay \By sub \Ax \Bx sub atan \Cy \By sub \Cx \Bx sub atan sub }%
\def\C{\By \Cy sub \Bx \Cx sub atan \Ay \Cy sub \Ax \Cx sub atan sub }%
\def\r{\b \c add \a sub \c \a add \b sub \a \b add \c sub mul mul
\a \b \c add add div sqrt 2 div }%
\def\R{\a \A sin div 2 div }%
\def\Area{\a \b \C sin mul mul 2 div }
\def\ha{\Area 2 mul \a div }
\def\hb{\Area 2 mul \b div }
\def\hc{\Area 2 mul \c div }
\def\HAx{\Ax \ha \Hy \Ay sub \Hx \Ax sub atan cos mul add }
\def\HAy{\Ay \ha \Hy \Ay sub \Hx \Ax sub atan sin mul add }
\def\HBx{\Bx \hb \Hy \By sub \Hx \Bx sub atan cos mul add }
\def\HBy{\By \hb \Hy \By sub \Hx \Bx sub atan sin mul add }
\def\HCx{\Cx \hc \Hy \Cy sub \Hx \Cx sub atan cos mul add }
\def\HCy{\Cy \hc \Hy \Cy sub \Hx \Cx sub atan sin mul add }
\def\baricentricX#1#2#3{\Ax #1 mul \Bx #2 mul \Cx #3 mul add add #1 #2
  #3 add add div }%
\def\baricentricY#1#2#3{\Ay #1 mul \By #2 mul \Cy #3 mul add add #1 #2
  #3 add add div }%
\def\trilinearX#1#2#3{\baricentricX{#1 \a mul }{#2 \b mul }{#3 \c mul }}%
\def\trilinearY#1#2#3{\baricentricY{#1 \a mul }{#2 \b mul }{#3 \c mul }}%
\def\Hx{\trilinearX{\B cos \C cos mul }{\C cos \A cos mul }{\A cos \B cos mul }}%
\def\Hy{\trilinearY{\B cos \C cos mul }{\C cos \A cos mul }{\A cos \B cos mul }}%
\psset{unit=1.0cm}%
\begin{pspicture}(0.5,-0.2)(9.5,6.5)
  \pnode(! \Ax \Ay){X_1}
  \pnode(! \Bx \By){X_2}
  \pnode(! \Cx \Cy){X_3}
  \pnode(! \Hx \Hy){H}
  \pnode(! \Bx \Ax sub \By \Ay sub){x_3}
  \pnode(! \Cx \Bx sub \Cy \By sub){x_1}
  \pnode(! \Ax \Cx sub \Ay \Cy sub){x_2}
  \ncline[linewidth=1.2pt]{*->}{X_1}{X_2}
  \Bput{$\vec{x}_3$}
  \ncline[linewidth=1.2pt]{*->}{X_2}{X_3}
  \Bput{$\vec{x}_1$}
  \ncline[linewidth=1.2pt]{*->}{X_3}{X_1}
  \Bput{$\vec{x}_2$}
  \uput[180](X_1){$\vec{X}_1$}
  \uput[-60](X_2){$\vec{X}_2$}
  \uput[90](X_3){$\vec{X}_3$}
  \psdots[linewidth=1.2pt](H)
  \psline[linestyle=dashed](X_1)(! \HAx \HAy)
  \psline[linestyle=dashed](X_2)(! \HBx \HBy)
  \psline[linestyle=dashed](X_3)(! \HCx \HCy)
  \uput[-65](H){$H$}
  \psarc(X_1){1}{(x_3)}{! \Hy \Ay sub \Hx \Ax sub atan}
  \uput{1.1}[! \By \Ay sub \Bx \Ax sub atan 
  \Hy \Ay sub \Hx \Ax sub atan add 2 div](X_1){$\frac{\pi}{2} - \phi_2$}
  \psarc(X_2){1}{(x_1)}{! \Hy \By sub \Hx \Bx sub atan}
  \uput{1.1}[! \Cy \By sub \Cx \Bx sub atan 
  \Hy \By sub \Hx \Bx sub atan add 2 div](X_2){$\frac{\pi}{2} - \phi_3$}
  \psarc(X_3){1}{(x_2)}{! \Hy \Cy sub \Hx \Cx sub atan}
  \uput{1.1}[! \Ay \Cy sub \Ax \Cx sub atan 
  \Hy \Cy sub \Hx \Cx sub atan add 2 div](X_3){$\frac{\pi}{2} - \phi_1$}
\end{pspicture}%
%

%% file: fig2.tex
\SpecialCoor
\def\Ax{1 }\def\Ay{0 }%
\def\Bx{9 }\def\By{2 }%
\def\Cx{2 }\def\Cy{6 }%
\def\a{\Bx \Cx sub dup mul \By \Cy sub dup mul add sqrt }%
\def\b{\Cx \Ax sub dup mul \Cy \Ay sub dup mul add sqrt }%
\def\c{\Ax \Bx sub dup mul \Ay \By sub dup mul add sqrt }%
\def\A{\Cy \Ay sub \Cx \Ax sub atan \By \Ay sub \Bx \Ax sub atan sub }%
\def\B{\Ay \By sub \Ax \Bx sub atan \Cy \By sub \Cx \Bx sub atan sub }%
\def\C{\By \Cy sub \Bx \Cx sub atan \Ay \Cy sub \Ax \Cx sub atan sub }%
\def\r{\b \c add \a sub \c \a add \b sub \a \b add \c sub mul mul
\a \b \c add add div sqrt 2 div }%
\def\baricentricX#1#2#3{\Ax #1 mul \Bx #2 mul \Cx #3 mul add add #1 #2
  #3 add add div }%
\def\baricentricY#1#2#3{\Ay #1 mul \By #2 mul \Cy #3 mul add add #1 #2
  #3 add add div }%
\def\trilinearX#1#2#3{\baricentricX{#1 \a mul }{#2 \b mul }{#3 \c mul }}%
\def\trilinearY#1#2#3{\baricentricY{#1 \a mul }{#2 \b mul }{#3 \c mul }}%
\def\Ix{\trilinearX{1}{1}{1}}%
\def\Iy{\trilinearY{1}{1}{1}}%
\psset{unit=1.0cm}%
\begin{pspicture}(0.5,-0.2)(9.5,6.5)
  \pnode(! \Ax \Ay){X_1}
  \pnode(! \Bx \By){X_2}
  \pnode(! \Cx \Cy){X_3}
  \pnode(! \Ix \Iy){In}
  \pnode(! \Bx \Ax sub \By \Ay sub){x_3}
  \pnode(! \Cx \Bx sub \Cy \By sub){x_1}
  \pnode(! \Ax \Cx sub \Ay \Cy sub){x_2}
  \ncline[linewidth=1.2pt]{*->}{X_1}{X_2}
  \Bput{$\vec{x}_3$}
  \ncline[linewidth=1.2pt]{*->}{X_2}{X_3}
  \Bput{$\vec{x}_1$}
  \ncline[linewidth=1.2pt]{*->}{X_3}{X_1}
  \Bput{$\vec{x}_2$}
  \uput[180](X_1){$\vec{X}_1$}
  \uput[-60](X_2){$\vec{X}_2$}
  \uput[90](X_3){$\vec{X}_3$}
  \ncline[linestyle=dashed]{-*}{X_1}{In}
  \ncline[linestyle=dashed]{-*}{X_2}{In}
  \ncline[linestyle=dashed]{-*}{X_3}{In}
  \psellipse(In)(! \r \r)
  \uput[-65](In){$I$}
  \psarc(X_1){1}{(x_3)}{! \Cy \Ay sub \Cx \Ax sub atan}
  \uput{1.1}[! \By \Ay sub \Bx \Ax sub atan 
  \Iy \Ay sub \Ix \Ax sub atan add 2 div](X_1){$\phi_1/2$}
  \psarc(X_2){1}{(x_1)}{! \Ay \By sub \Ax \Bx sub atan}
  \uput{1.1}[! \Cy \By sub \Cx \Bx sub atan 
  \Iy \By sub \Ix \Bx sub atan add 2 div](X_2){$\phi_2/2$}
  \psarc(X_3){1}{(x_2)}{! \By \Cy sub \Bx \Cx sub atan}
  \uput{1.1}[! \Ay \Cy sub \Ax \Cx sub atan 
  \Iy \Cy sub \Ix \Cx sub atan add 2 div](X_3){$\phi_3/2$}
\end{pspicture}%
%

%% file: fig5.tex
\SpecialCoor
\def\Ax{1 }\def\Ay{0 }%
\def\Bx{9 }\def\By{2 }%
\def\Cx{2 }\def\Cy{6 }%
\def\a{\Bx \Cx sub dup mul \By \Cy sub dup mul add sqrt }%
\def\b{\Cx \Ax sub dup mul \Cy \Ay sub dup mul add sqrt }%
\def\c{\Ax \Bx sub dup mul \Ay \By sub dup mul add sqrt }%
\def\A{\Cy \Ay sub \Cx \Ax sub atan \By \Ay sub \Bx \Ax sub atan sub }%
\def\B{\Ay \By sub \Ax \Bx sub atan \Cy \By sub \Cx \Bx sub atan sub }%
\def\C{\By \Cy sub \Bx \Cx sub atan \Ay \Cy sub \Ax \Cx sub atan sub }%
\def\r{\b \c add \a sub \c \a add \b sub \a \b add \c sub mul mul
\a \b \c add add div sqrt 2 div }%
\def\R{\a \A sin div 2 div }%
\def\Area{\a \b \C sin mul mul 2 div }
\def\ha{\Area 2 mul \a div }
\def\hb{\Area 2 mul \b div }
\def\hc{\Area 2 mul \c div }
\def\BarAx{\Bx \Cx add 2 div }
\def\BarAy{\By \Cy add 2 div }
\def\BarBx{\Cx \Ax add 2 div }
\def\BarBy{\Cy \Ay add 2 div }
\def\BarCx{\Ax \Bx add 2 div }
\def\BarCy{\Ay \By add 2 div }
\def\baricentricX#1#2#3{\Ax #1 mul \Bx #2 mul \Cx #3 mul add add #1 #2
  #3 add add div }%
\def\baricentricY#1#2#3{\Ay #1 mul \By #2 mul \Cy #3 mul add add #1 #2
  #3 add add div }%
\def\trilinearX#1#2#3{\baricentricX{#1 \a mul }{#2 \b mul }{#3 \c mul }}%
\def\trilinearY#1#2#3{\baricentricY{#1 \a mul }{#2 \b mul }{#3 \c mul }}%
\def\Barx{\baricentricX{1}{1}{1}}%
\def\Bary{\baricentricY{1}{1}{1}}%
\psset{unit=1.0cm}%
\begin{pspicture}(0.5,-0.2)(9.5,6.5)
  \pnode(! \Ax \Ay){X_1}
  \pnode(! \Bx \By){X_2}
  \pnode(! \Cx \Cy){X_3}
  \pnode(! \Barx \Bary){Bar}
  \pnode(! \Bx \Ax sub \By \Ay sub){x_3}
  \pnode(! \Cx \Bx sub \Cy \By sub){x_1}
  \pnode(! \Ax \Cx sub \Ay \Cy sub){x_2}
  \ncline[linewidth=1.2pt]{*->}{X_1}{X_2}
  \Bput{$\vec{x}_3$}
  \ncline[linewidth=1.2pt]{*->}{X_2}{X_3}
  \Bput{$\vec{x}_1$}
  \ncline[linewidth=1.2pt]{*->}{X_3}{X_1}
  \Bput{$\vec{x}_2$}
  \uput[180](X_1){$\vec{X}_1$}
  \uput[-60](X_2){$\vec{X}_2$}
  \uput[90](X_3){$\vec{X}_3$}
  \psdots[linewidth=1.2pt](Bar)
  \pcline[linestyle=dashed](X_1)(! \BarAx \BarAy)
  \lput*{0}{$h_1$}
  \pcline[linestyle=dashed](X_2)(! \BarBx \BarBy)
  \lput*{0}{$h_2$}
  \pcline[linestyle=dashed](X_3)(! \BarCx \BarCy)
  \lput*{0}{$h_3$}
  \uput[90](Bar){$B$}
  \psarc(X_1){1}{(x_3)}{! \Cy \Ay sub \Cx \Ax sub atan}
  \psarc(! \BarAx \BarAy){1}{! \Cy \BarAy sub \Cx \BarAx sub atan}%
  {! \Ay \BarAy sub \Ax \BarAx sub atan}
  \uput{1.1}[! \Cy \BarAy sub \Cx \BarAx sub atan \Ay \BarAy sub \Ax
  \BarAx sub atan add 2 div](! \BarAx \BarAy){$\psi_1$}
  \psarc(X_2){1}{(x_1)}{! \Ay \By sub \Ax \Bx sub atan}
  \psarc(! \BarBx \BarBy){1}{! \Ay \BarBy sub \Ax \BarBx sub atan}%
  {! \By \BarBy sub \Bx \BarBx sub atan}
  \uput{1.1}[! \Ay \BarBy sub \Ax \BarBx sub atan \By \BarBy sub \Bx
  \BarBx sub atan add 2 div](! \BarBx \BarBy){$\psi_2$}
  \psarc(X_3){1}{(x_2)}{! \By \Cy sub \Bx \Cx sub atan}
  \psarc(! \BarCx \BarCy){1}{! \By \BarCy sub \Bx \BarCx sub atan}%
  {! \Cy \BarCy sub \Cx \BarCx sub atan}
  \uput{1.1}[! \By \BarCy sub \Bx \BarCx sub atan \Cy \BarCy sub \Cx
  \BarCx sub atan add 2 div](! \BarCx \BarCy){$\psi_3$}
\end{pspicture}%
%

%% file: fig3.tex
\SpecialCoor
\def\Ax{1 }\def\Ay{0 }%
\def\Bx{9 }\def\By{2 }%
\def\Cx{2 }\def\Cy{6 }%
\def\a{\Bx \Cx sub dup mul \By \Cy sub dup mul add sqrt }%
\def\b{\Cx \Ax sub dup mul \Cy \Ay sub dup mul add sqrt }%
\def\c{\Ax \Bx sub dup mul \Ay \By sub dup mul add sqrt }%
\def\A{\Cy \Ay sub \Cx \Ax sub atan \By \Ay sub \Bx \Ax sub atan sub }%
\def\B{\Ay \By sub \Ax \Bx sub atan \Cy \By sub \Cx \Bx sub atan sub }%
\def\C{\By \Cy sub \Bx \Cx sub atan \Ay \Cy sub \Ax \Cx sub atan sub }%
\def\r{\b \c add \a sub \c \a add \b sub \a \b add \c sub mul mul
\a \b \c add add div sqrt 2 div }%
\def\R{\a \A sin div 2 div }%
\def\baricentricX#1#2#3{\Ax #1 mul \Bx #2 mul \Cx #3 mul add add #1 #2
  #3 add add div }%
\def\baricentricY#1#2#3{\Ay #1 mul \By #2 mul \Cy #3 mul add add #1 #2
  #3 add add div }%
\def\trilinearX#1#2#3{\baricentricX{#1 \a mul }{#2 \b mul }{#3 \c mul }}%
\def\trilinearY#1#2#3{\baricentricY{#1 \a mul }{#2 \b mul }{#3 \c mul }}%
\def\Ox{\trilinearX{\A cos}{\B cos }{\C cos }}%
\def\Oy{\trilinearY{\A cos}{\B cos }{\C cos }}%
\psset{unit=1.0cm}%
\begin{pspicture}(0,-2.2)(9.7,7)
  \pnode(! \Ax \Ay){X_1}
  \pnode(! \Bx \By){X_2}
  \pnode(! \Cx \Cy){X_3}
  \pnode(! \Ox \Oy){Out}
  \pnode(! \Bx \Ax sub \By \Ay sub){x_3}
  \pnode(! \Cx \Bx sub \Cy \By sub){x_1}
  \pnode(! \Ax \Cx sub \Ay \Cy sub){x_2}
  \ncline[linewidth=1.2pt]{*->}{X_1}{X_2}
  \Bput{$\vec{x}_3$}
  \ncline[linewidth=1.2pt]{*->}{X_2}{X_3}
  \Bput{$\vec{x}_1$}
  \ncline[linewidth=1.2pt]{*->}{X_3}{X_1}
  \Bput{$\vec{x}_2$}
  \uput[180](X_1){$\vec{X}_1$}
  \uput[-60](X_2){$\vec{X}_2$}
  \uput[90](X_3){$\vec{X}_3$}
  \ncline[linestyle=dashed]{-*}{X_1}{Out}
  \ncline[linestyle=dashed]{-*}{X_2}{Out}
  \ncline[linestyle=dashed]{-*}{X_3}{Out}
  \psellipse(Out)(! \R \R)
  \uput[-65](Out){$O$}
  \psarc(X_1){1}{(x_3)}{! \Oy \Ay sub \Ox \Ax sub atan}
  \uput{1.1}[! \By \Ay sub \Bx \Ax sub atan 
  \Oy \Ay sub \Ox \Ax sub atan add 2 div](X_1){$\frac{\pi}{2} - \phi_3$}
  \psarc(X_2){1}{(x_1)}{! \Oy \By sub \Ox \Bx sub atan}
  \uput{1.1}[! \Cy \By sub \Cx \Bx sub atan 
  \Oy \By sub \Ox \Bx sub atan add 2 div](X_2){$\frac{\pi}{2} - \phi_1$}
  \psarc(X_3){1}{(x_2)}{! \Oy \Cy sub \Ox \Cx sub atan}
  \uput{1.1}[! \Ay \Cy sub \Ax \Cx sub atan 
  \Oy \Cy sub \Ox \Cx sub atan add 2 div](X_3){$\frac{\pi}{2} - \phi_2$}
  \psarc(Out){1}{! \Cy \Oy sub \Cx \Ox sub atan}%
  {! \Ay \Oy sub \Ax \Ox sub atan}
  \uput{1.1}[! \Cy \Oy sub \Cx \Ox sub atan \Ay \Oy sub \Ax \Ox sub
  atan add 2 div](Out){$2 \phi_2$}
\end{pspicture}%
%

%% file: fig6.tex
\SpecialCoor
\def\Ax{1 }\def\Ay{0 }%
\def\Bx{9 }\def\By{2 }%
\def\Cx{2 }\def\Cy{6 }%
\def\a{\Bx \Cx sub dup mul \By \Cy sub dup mul add sqrt }%
\def\b{\Cx \Ax sub dup mul \Cy \Ay sub dup mul add sqrt }%
\def\c{\Ax \Bx sub dup mul \Ay \By sub dup mul add sqrt }%
\def\A{\Cy \Ay sub \Cx \Ax sub atan \By \Ay sub \Bx \Ax sub atan sub }%
\def\B{\Ay \By sub \Ax \Bx sub atan \Cy \By sub \Cx \Bx sub atan sub }%
\def\C{\By \Cy sub \Bx \Cx sub atan \Ay \Cy sub \Ax \Cx sub atan sub }%
\def\r{\b \c add \a sub \c \a add \b sub \a \b add \c sub mul mul
\a \b \c add add div sqrt 2 div }%
\def\R{\a \A sin div 2 div }%
\def\Area{\a \b \C sin mul mul 2 div }
\def\baricentricX#1#2#3{\Ax #1 mul \Bx #2 mul \Cx #3 mul add add #1 #2
  #3 add add div }%
\def\baricentricY#1#2#3{\Ay #1 mul \By #2 mul \Cy #3 mul add add #1 #2
  #3 add add div }%
\def\trilinearX#1#2#3{\baricentricX{#1 \a mul }{#2 \b mul }{#3 \c mul }}%
\def\trilinearY#1#2#3{\baricentricY{#1 \a mul }{#2 \b mul }{#3 \c mul
  }}%
\def\Ix{\trilinearX{1}{1}{1}}%
\def\Iy{\trilinearY{1}{1}{1}}%
\def\JAx{\trilinearX{-1}{1}{1}}%
\def\JAy{\trilinearY{-1}{1}{1}}%
\def\JBx{\trilinearX{1}{-1}{1}}%
\def\JBy{\trilinearY{1}{-1}{1}}%
\def\JCx{\trilinearX{1}{1}{-1}}%
\def\JCy{\trilinearY{1}{1}{-1}}%
\def\rA{\Area 2 mul \b \c add \a sub div }
\def\rB{\Area 2 mul \c \a add \b sub div }
\def\rC{\Area 2 mul \a \b add \c sub div }
\psset{unit=0.7cm}%
\begin{pspicture}(-3.7,-6.7)(10.8,10.6)
  \pnode(! \Ax \Ay){X_1}
  \pnode(! \Bx \By){X_2}
  \pnode(! \Cx \Cy){X_3}
  \pnode(! \Ix \Iy){In}
  \pnode(! \JAx \JAy){J_1}
  \pnode(! \JBx \JBy){J_2}
  \pnode(! \JCx \JCy){J_3}
  \pnode(! \Bx \Ax sub \By \Ay sub){x_3}
  \pnode(! \Cx \Bx sub \Cy \By sub){x_1}
  \pnode(! \Ax \Cx sub \Ay \Cy sub){x_2}
  \ncline[linewidth=1.2pt]{*->}{X_1}{X_2}
  \Bput{$\vec{x}_3$}
  \ncline[linewidth=1.2pt]{*->}{X_2}{X_3}
  \Bput{$\vec{x}_1$}
  \ncline[linewidth=1.2pt]{*->}{X_3}{X_1}
  \Bput{$\vec{x}_2$}
  \uput[180](X_1){$\vec{X}_1$}
  \uput[-60](X_2){$\vec{X}_2$}
  \uput[90](X_3){$\vec{X}_3$}
  \psset{linestyle=dashed}
  \psdots(In)
  \uput[90](In){$I$}
  \ncline{-*}{X_1}{J_1}
  \ncline{-*}{X_2}{J_1}
  \ncline{-*}{X_3}{J_1}
  \uput[0](J_1){$J_1$}
  \ncline{-*}{X_1}{J_2}
  \ncline{-*}{X_2}{J_2}
  \ncline{-*}{X_3}{J_2}
  \uput[120](J_2){$J_2$}
  \ncline{-*}{X_1}{J_3}
  \ncline{-*}{X_2}{J_3}
  \ncline{-*}{X_3}{J_3}
  \uput[0](J_3){$J_3$}
  \begin{psclip}{\psframe[linestyle=none](-3.7,-6.7)(10.8,10.6)}
    \psset{linestyle=dotted}
    \psline([angle=! \By \Ay sub \Bx \Ax sub atan,nodesep=-100]X_1)%
    ([angle=! \By \Ay sub \Bx \Ax sub atan,nodesep=100]X_1)
    \psline([angle=! \Cy \By sub \Cx \Bx sub atan,nodesep=-100]X_2)%
    ([angle=! \Cy \By sub \Cx \Bx sub atan,nodesep=100]X_2)
    \psline([angle=! \Ay \Cy sub \Ax \Cx sub atan,nodesep=-100]X_3)%
    ([angle=! \Ay \Cy sub \Ax \Cx sub atan,nodesep=100]X_3)
    \psset{linestyle=dotted}
    \psellipse(J_1)(! \rA \rA)
    \psellipse(J_2)(! \rB \rB)
    \psellipse(J_3)(! \rC \rC)
  \end{psclip}
\end{pspicture}%
%